\documentclass[11pt,twoside]{article}

\usepackage{asp2006}
\usepackage{epsf}
\usepackage{psfig}
\usepackage{lscape}

\newcommand{\xmm}{XMM--{\it Newton}}
\newcommand{\flux}{erg cm$^{-2}$ s$^{-1}$}

\begin{document}
\title{X-ray source counts in the XMM-COSMOS survey}   
\author{
N.~Cappelluti\altaffilmark{1},
G.~Hasinger\altaffilmark{1},
M.~Brusa\altaffilmark{1},
A.~Comastri\altaffilmark{2},
G.~Zamorani\altaffilmark{2},
H.~B\"ohringer\altaffilmark{1},
H.~Brunner\altaffilmark{1},
F.~Civano\altaffilmark{2,4},
A.~Finoguenov\altaffilmark{1},
F.~Fiore\altaffilmark{5}, 
R.~Gilli\altaffilmark{2},
R.~E. Griffiths\altaffilmark{3},
V.~Mainieri\altaffilmark{1},
I.~Matute\altaffilmark{1},
T.~Miyaji\altaffilmark{3},
J.~Silverman\altaffilmark{1}
}
\altaffiltext{1}{Max Planck Institut f\"ur Extraterrestrische Physik,~D-85478 Garching, Germany}
\altaffiltext{2}{INAF-Osservatorio Astronomico di Bologna, via Ranzani 1, I-40127 Bologna, Italy}
\altaffiltext{3}{Department of Physics, Carnegie Mellon University, 5000 Forbes Avenue, Pittsburgh, PA 15213}
\altaffiltext{4}{Dipartimento di Astronomia, Universit\`a di Bologna,
    via Ranzani 1, I--40127 Bologna, Italy}
\altaffiltext{5}{INAF-Osservatorio astronomico di Roma, Via Frascati 33, I-00044 Monteporzio Catone, Italy}
\begin{abstract} 
We present the analysis of the source counts in the XMM-COSMOS 
survey using data of the first year of XMM-{\em Newton} observations. 
The survey covers $\sim$2 deg$^{2}$ within the region of sky bounded by
$9^h57.5^m<R.A.<10^h03.5^m$; $1^d27.5^m<DEC<2^d57.5^m$ with a total net
integration time of 504 ks. Using a maximum likelihood algorithm 
we detected a total of 1390 sources at least in one band.
Using Monte Carlo simulations to estimate the sky coverage we produced
the logN-logS relations. These relations
have been then derived in the 0.5--2 keV, 2--10 keV and 5--10 keV energy bands,
down to  flux
limits of  7.2$\times$10$^{-16}$ erg cm$^{-2}$ s$^{-1}$, 4.0$\times$10$^{-15}$
erg cm$^{-2}$ s$^{-1}$ and  9.7$\times$10$^{-15}$ erg cm$^{-2}$ s$^{-1}$, respec
tively. These relations have been compared to previous X-ray survey and to the most recent
X-ray background model finding an excellent agreement. 
The slightly different normalizations observed in t
he source counts of COSMOS and previous surveys can be largely explained as a co
mbination of low counting statistics and cosmic variance introduced by the large
scale structure.
\end{abstract}
\vspace{-1cm}
\section{Introduction}
A solid knowledge of the  X-ray source counts is fundamental to fully understand
the AGNs population of the X-ray background (XRB). 
At present the two deepest X--ray surveys, the {\it Chandra} Deep
Field North \citep[CDFN;][]{cap:bau04} and  {\it Chandra} Deep Field South \citep[CDFS;][]{cap:gia01},
 have extended the sensitivity by about two orders of magnitude 
in all bands with respect to previous surveys \citep{cap:has93,cap:ue99,cap:gio00}, detecting
a large number  of faint X--ray sources.
However, deep  pencil beam surveys are limited by the area which can
be covered to very faint fluxes (typically of the order of 0.1 deg$^2$) and
suffer from significant field to field variance.
To ovecome this problem, shallower surveys over  larger
areas have been undertaken in the last few years with both {\it Chandra}
(e.g. the 9 deg$^{2}$ Bootes survey \citep{cap:mu05}, the Extended Groth strip EGS \
\citep{cap:nan05},  the Extended {\it Chandra} Deep Field South E-CDFS, \citep{cap:leh06,vir06} and th
e Champ \citep{cap:gre04,cap:kim04}) and \xmm ~(e.g. the HELLAS2XMM survey \citep{cap:fio03}, 
the \xmm ~ BSS \citep{cap:dc04}  and the  ELAIS S1 survey \citep{cap:puc06} ).
In this context the \xmm~ wide field
survey in the COSMOS field \citep{cap:scoville}, hereinafter XMM--COSMOS \citep{cap:has06},
 has been conceived and designed to maximize
the sensitivity and survey area product, and  is expected to provide
a major step forward toward a complete characterization of
the physical properties of X--ray emitting Super Massive Black Holes (SMBHs).
In this work we concentrate on the first year of XMM-{\em Newton} observations (AO3)
of the COSMOS field \citep{cap:has06}. For  more details see Cappelluti et~al. (2007).
\section{The X-ray logN-logS}
  The source detection was performed using a maximum likelihood algorithm provided with the
  \xmm~ Standard Analysis System (SAS) in the 0.5--2 keV, 2--10 keV and 5--10 keV bands. 
   A total of A total of 1281, 724 and 186 point-like sources were detected in the three
 bands down to  limiting fluxes of 7.2$\times$10$^{-16}$ erg cm$^{-2}$
 s$^{-1}$, 4.7$\times$10$^{-15}$ erg cm$^{-2}$ s$^{-1}$ and
 9.7$\times$10$^{-15}$ erg cm$^{-2}$ s$^{-1}$, respectively.
 Using Monte Carlo simulation to determine the sky coverage, 
source number counts can be easily computed using the following equation:
$N(>S)=\sum_{i=1}^{\it N_S} \frac{1}{\Omega_{i}} deg^{-2}$,
where {\it $N_S$} is the total number of detected sources in the field
with fluxes greater than $S$ and $\Omega_{i}$ is the sky coverage
associated with the flux of the i$^{th}$ source. 
The cumulative number counts, normalized to the Euclidean slope (multiplied by
S$^{1.5}$), are shown in 
Figure 1, 
for the 0.5--2 keV, 2--10 keV and 5--10 keV energy ranges, respectively.
We performed a broken power-law  maximum likelihood fit  
to the unbinned differential counts in the 0.5--2 keV and 2--10 keV energy bands.
In the 0.5--2 keV energy band
the best fit parameters are $\alpha_{1}$=2.60$^{+0.15}_{-0.18}$, $\alpha_{2}$=1.65$\pm{0.05}$,
$S_{b}$=1.55$^{+0.28}_{-0.24}$ $\times$10$^{-14}$ \flux ~ and $A$=123. 
Translating this value of the normalization to that for the cumulative 
distribution at 2$\times$10$^{-15}$ \flux ~, which is usually used in the literature
for {\it Chandra} surveys, we obtain $A_{15}\sim$450 which is fully consistent with 
most of previous works where a fit result is presented.
In the 2--10 keV band the best fit parameters are $\alpha_{1}$=2.43$\pm{0.10}$, $\alpha_{2}$=1.59$\pm{0.33}$,
$S_{b}$=1.02$^{+0.25}_{-0.19}$ $\times$10$^{-14}$ \flux ~ and $A$=266. The latest value translates into  
$A_{15}\sim$1250. Also in this band, our results are in  agreement  with previous surveys within 1$\sigma$.
In the 5--10 keV energy bands, where
the differential counts do not show any  evidence for a break in the sampled 
flux range, we assumed a single power-law model for which the best fit parameters are found to be $A$=102 and 
$\alpha_{1}$=2.36$\pm{0.1}$. 
In the 0.5--2 keV band we measure a  contribution of the sources to the XRB which corresponds to
 a normalization at 1 keV of 
7.2$\pm{1.1}$ keV cm$^{-2}$ s$^{-1}$ keV$^{-1}$. The corresponding values in the 
2--10 keV and 5--10 keV bands are  4.7$\pm{1.1}$ keV cm$^{-2}$ s$^{-1}$ keV$^{-1}$  and 
  2.6 $\pm{1.4}$ keV cm$^{-2}$ s$^{-1}$ keV$^{-1}$. 
Therefore XMM-COSMOS resolves by itself $\sim$65$\%$, $\sim$40$\%$ and $\sim$22$\%$ of the XRB into
discrete sources in the  0.5--2 keV, 2--10 keV and 5--10 keV energy bands, respectively. 
In Figure 1 we compared our logN-logS
 to those predicted by the recent XRB model of \citet{cap:gil06}. 
\begin{figure}[!t]
\begin{center}
\hspace{-0cm}
\psfig{file=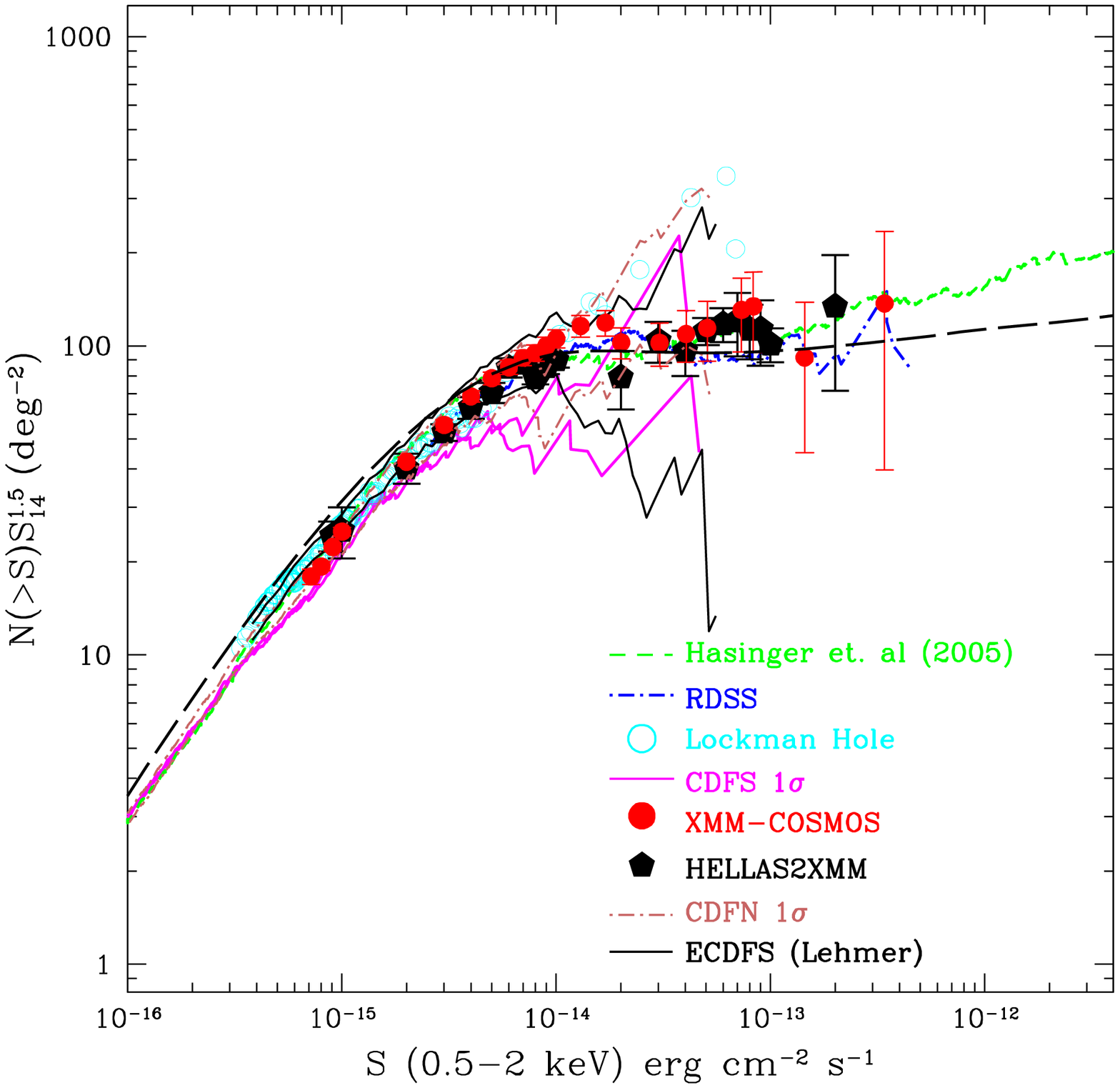,width=6.5cm,height=5.5cm,angle=0}
\psfig{file=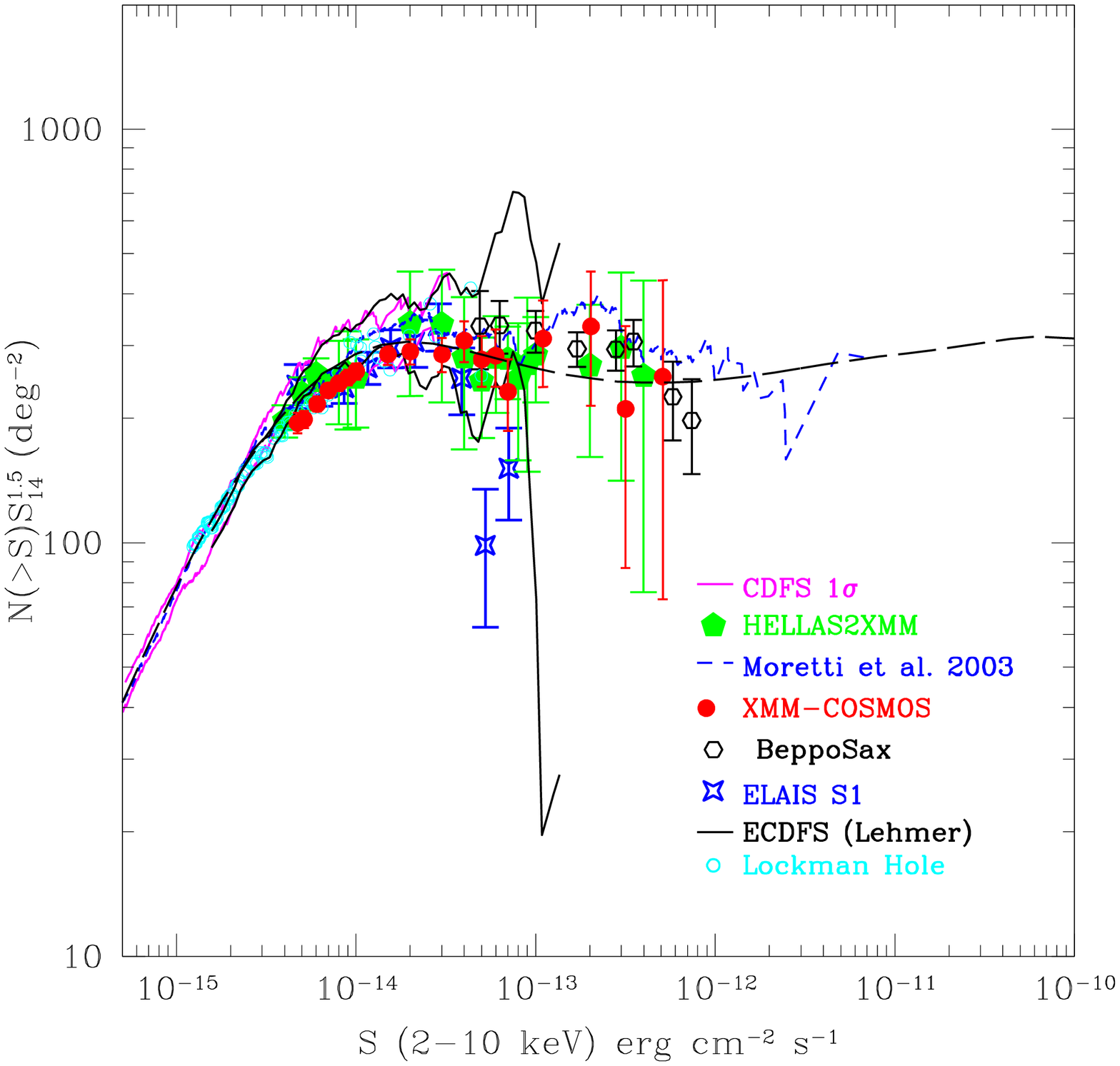,width=6.5cm,height=5.5cm,angle=0}
\psfig{file=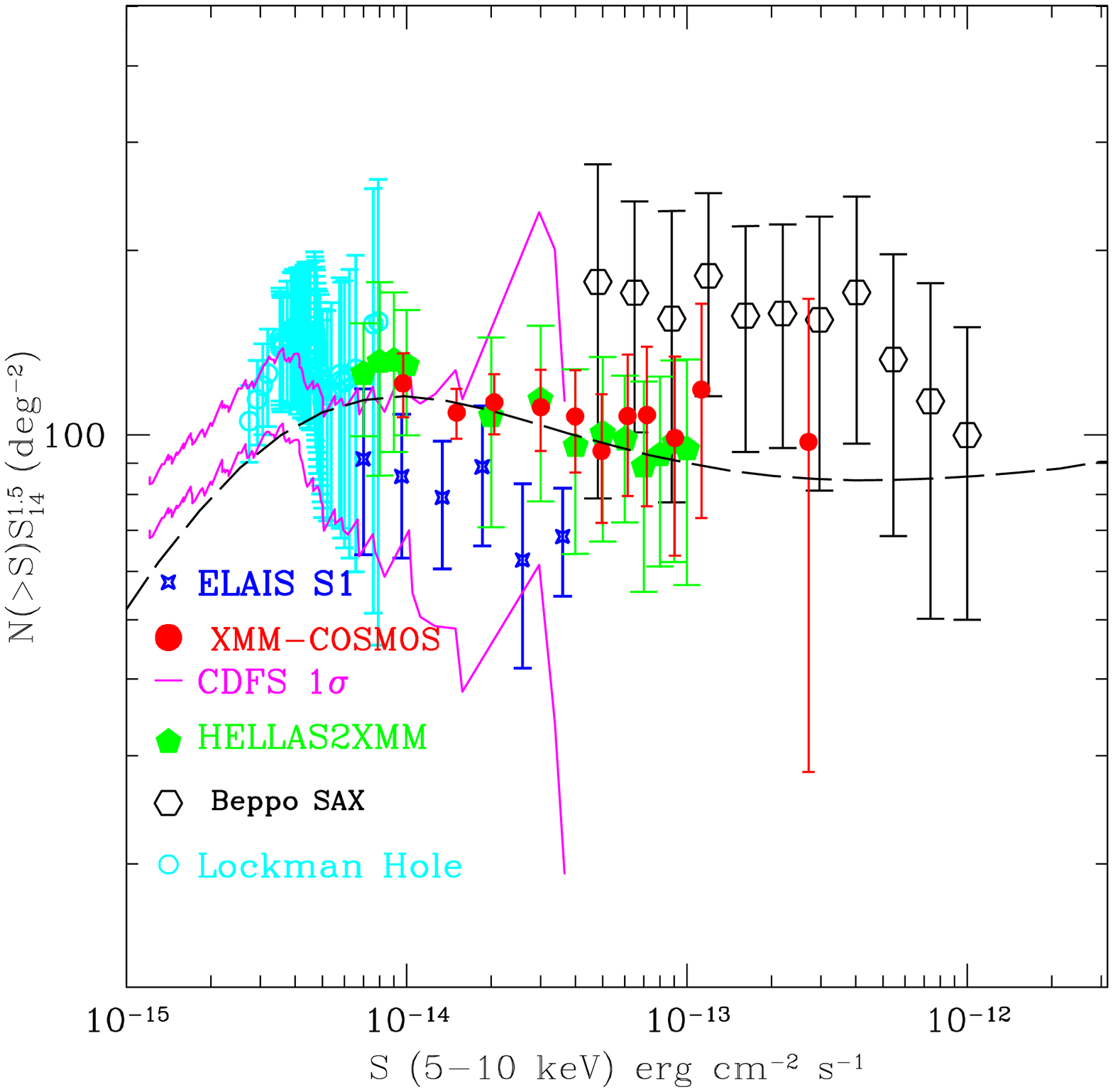,width=6.5cm,height=5.5cm,angle=0}
\psfig{file=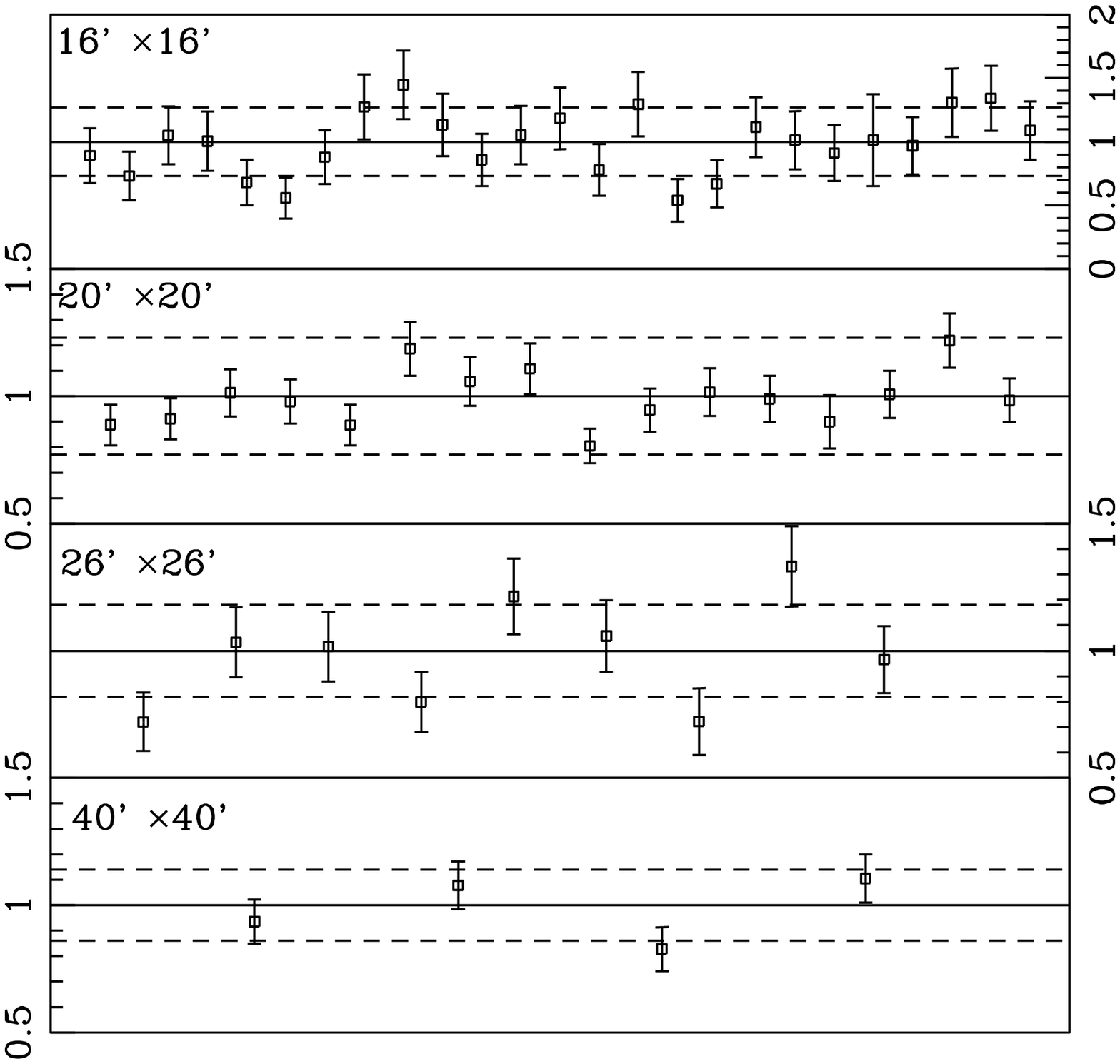,width=6.5cm,height=5.5cm,angle=0}
\caption{ The X-ray logN-logS compared to the survey listed in the labels in the 0.5--2 keV, 2--10 keV and 5--10 keV energy bands, respectively $Top~Left,~Top~Right~and~Bottom~Left~Panels$. The $black-dashed$ line represents the prediction of \citet{cap:gil06}. $Bottom~Right~Panel:$ The  counts in cell fluctuations  within the XMM-COSMOS field on different angular scales. The dashed lines represent the 1$\sigma$ expected fluctuation.}
\end{center}
\end{figure}

The amplitude of  source counts distributions  varies
significantly among different surveys \citep[see e.g.][and references therein]{cap:ya03,cap:cap05}.
This  "sample variance", can be explained and predicted as a combination of Poissonian variations and effects
due to the clustering of sources \citep{cap:pee,cap:ya03}.
In order to determine whether  the differences observed in  the source counts of different
 surveys could arise  from the clustering of X-ray sources, 
 we estimated  the amplitude of the fluctuations from our data, 
 by producing subsamples of our survey with areas  comparable 
 to those of. e.g., {\it Chandra}  surveys. \\ 
The XMM-COSMOS field and the Monte Carlo sample fields  of Section 4 were divided in 4,9,16 
and 25 square boxes.
Making use of the 0.5--2 keV energy band data, we  computed for each subfield, the  ratio of
 the number of real sources to the number of random  source. Both the random and the real sample 
 were cut to a flux limit of 5$\times$10$^{-15}$ \flux.
In the lower right panel of Figure 1 we plot the ratio of the data to the random sample as a function
 of the size of the cells under investigation.
 The predicted fractional standard deviations are 0.13, 0.19, 0.23 and 0.28
  on scales of  0.44 deg$^{2}$, 0.19 deg$^{2}$, 0.11 deg$^{2}$  and 0.07 deg$^{2}$, respectively.  
The measure fractional standard deviations are 0.09, 0.20, 0.21 and 0.24 on the same scales, respectively. 
The ratio of the clusting to the Poissonian variance is expected to scale as   
$\sigma_{cl}/ \sigma_{p}\propto  \mathcal{N}^{0.5} \theta_{0}^{(\gamma-1)/2} a^{(3-\gamma)/2}$. 
We therefore conclude
that at this scales and at fluxes sampled here the major contribution to source counts fluctuations is due to
Poissonian noise.
This analysis is at least qualitatively consistent with Figure 2, which shows a 
 significantly larger dispersion in the data from different surveys in the hard band than
 in the soft band. Moreover, the results here discussed are also consistent with the
 observed fluctuations in the deep Chandra fields (see, for example,  Bauer et al. 2004).
 Large area, moderately deep surveys like XMM-COSMOS are needed to overcome the problem 
of low counting statistics, typical of deep pencil beam surveys, and, at the same time, 
to provide a robust estimate of the effect of large scale structure on observed source 
counts.
\acknowledgements
This work is based on observations
   obtained with \xmm~, an ESA science mission with
instruments and contributions directly funded by ESA Member States and NASA;
also based on data collected at the Canada-France-Hawaii Telescope
operated by the National Research Council of Canada,  the Centre National de
la Recherche Scientifique de France and the University of Hawaii.


\begin{thebibliography}{}

\bibitem[Bauer et al.(2004)]{cap:bau04} Bauer, F.~E., Alexander, 
D.~M., Brandt, W.~N., Schneider, D.~P., Treister, E., Hornschemeier, A.~E., 
\& Garmire, G.~P.\ 2004, \aj, 128, 2048 
\bibitem[Cappelluti et~al. (2005){}{}{}]{cap:cap05} Cappelluti, N.,
Cappi, M., Dadina, M., Malaguti, G., Branchesi, M., D'Elia, V., \& Palumbo,
G.~G.~C.\ 2005, \aap, 430, 39
\bibitem[Cappelluti et~al. (2007){}{}{}]{cap:cap07} Cappelluti, N., et~al\ 2007, \apjs in press, 
 ArXiv Astrophysics e-prints, arXiv:astro-ph/0701196 
\bibitem[Della Ceca et al.(2004)]{cap:dc04} Della Ceca, R., et 
al.\ 2004, \aap, 428, 383 
\bibitem[Fiore et al.(2003)]{cap:fio03} Fiore, F., et al.\ 2003, 
\aap, 409, 79   
\bibitem[Giacconi et al.(2001)]{cap:gia01} Giacconi, R., et al.\ 
2001, \apj, 551, 624 
\bibitem[Gilli, Comastri \& Hasinger (2006)]{cap:gil06} Gilli, R., Comastri, A. \&
  Hasinger, G. 2006, A\&A, in press, astro-ph/0610939 
\bibitem[Giommi et al.(2000)]{cap:gio00} Giommi, P., Perri, M., 
\& Fiore, F.\ 2000, \aap, 362, 799  
\bibitem[Green et al.(2004)]{cap:gre04} Green, P.~J., et al.\ 
2004, \apjs, 150, 43 
\bibitem[Hasinger et al.(1993)]{cap:has93} Hasinger, G., Burg, 
R., Giacconi, R., Hartner, G., Schmidt, M., Trumper, J., \& Zamorani, G.\ 
1993, \aap, 275, 1 
\bibitem[Hasinger et al.(2006)]{cap:has06} Hasinger, G. et al.  2006, \apjs in press,
ArXiv Astrophysics e-prints, arXiv:astro-ph/0612311 
\bibitem[Lehmer et al.(2005)]{cap:leh06} Lehmer, B.~D., et al.\ 
2005, \apjs, 161, 21
\bibitem[Kim et al.(2004)]{cap:kim04} Kim, D.-W., et al.\ 2004, 
\apj, 600, 59 
\bibitem[Murray et al.(2005)]{cap:mu05} Murray, S.~S., et al.\ 
2005, \apjs, 161, 1
\bibitem[Nandra et al.(2005)]{cap:nan05} Nandra, K., et al.\ 
2005, \mnras, 356, 568
\bibitem[Peebles (1980)]{cap:pee}Peebles, P. J. E. 1980, The large-scale structure of the universe (Princeton, N.J., Princeton University Press)
\bibitem[Puccetti et al.(2006)]{cap:puc06} Puccetti  et al.\ 
2006, \aap, 457, 501
\bibitem[Scoville et al.(2006))]{cap:scoville} Scoville, N.Z. et al., 2007, \apjs in press,
2006, ArXiv Astrophysics e-prints, arXiv:astro-ph/0612305 
\bibitem[Ueda et al.(1999)]{cap:ue99} Ueda, Y., et al.\ 1999, 
\apj, 518, 656 
\bibitem[Virani et al.(2006)]{vir06} Virani, S.~N., Treister, 
E., Urry, C.~M., \& Gawiser, E.\ 2006, \aj, 131, 2373 
\bibitem[Yang et al.(2003)]{cap:ya03} Yang, Y., Mushotzky, 
R.~F., Barger, A.~J., Cowie, L.~L., Sanders, D.~B., \& Steffen, A.~T.\ 
2003, \apjl, 585, L85 
\end{thebibliography}
\end{document}